Holger Arndt# Economic and Financial Learning with Artificial Intelligence: A Mixed-Methods Study on ChatGPT.

**Highlights**

- Participants exhibited a significant positive shift in their perceptions of ChatGPT postexposure.
- Effective prompting was identified as a central challenge, with many underestimating its importance.
- The study introduced the concept of 'AI-learning-competence', hinting at a broader educational paradigm shift.

**Abstract**

In the evolving landscape of digital education, chatbots have emerged as potential game-changers, promising personalized and adaptive learning experiences. This research undertook an in-depth exploration of ChatGPT's potential as an educational tool, focusing on user perceptions, experiences and learning outcomes. Through a mixed-methods approach, a diverse group of 102 participants engaged with ChatGPT, providing insights pre- and postinteraction. The study reveals a notable positive shift in perceptions after exposure, underscoring the efficacy of ChatGPT. However, challenges such as prompting effectiveness and information accuracy emerged as pivotal concerns. Introducing the concept of 'AI-learning-competence', this study lays the groundwork for future research, emphasizing the need for formal training and pedagogical integration of AI tools.

# 1. Introduction

In the era of relentless digital evolution, artificial intelligence (AI) has seamlessly woven its way into various sectors, with education standing out as a domain ready for transformation. Among the many facets of AI, chatbots, powered by leaps in natural language processing, have emerged as a beacon for revolutionizing learning experiences. These digital dialog partners, with their capacity for instantaneous feedback, and personalization, ushering in a new dawn in contemporary pedagogy.

ChatGPT, developed by OpenAI, has particularly attracted the attention of the academic community thanks to its sophisticated conversational skills. ChatGPT is provided with the potential to act as a catalyst for knowledge, since everyone has easy access to it. Therefore, it could act as a linchpin in the realm of personalized learning. While the technological merits of ChatGPT and its ilk are hard to dispute, a deeper exploration into their efficacy in real-world educational contexts and the perceptions they engender is crucial.

In this study, the focus is on unraveling the multifaceted experience of employing ChatGPT as a learning ally. By utilizing a mixed-methods research design, the objective becomes twofold: to quantify the tangible impacts of ChatGPT on learning trajectories and to delve into the lived experiences and perceptions of learners. This comprehensive inquiry seeks to shed light on the opportunities, challenges, and potential of integrating ChatGPT into the educational landscape.

The subsequent sections are organized to provide readers with a comprehensive understanding of the study and its implications. The article begins with a literature review, tracing the evolution of AI in education and the emergence of chatbots as learning facilitators. Following this, the research questions are presented. Then the methodology section elucidates the research design, participant selection, and data collection processes. The results section presents a detailed analysis of both quantitative outcomes and qualitative insights. This is succeeded by discussions that interpret the findings, delving into their implications and situating them within broader pedagogical contexts. The article concludes with a reflection on the study's limitations and charts out potential avenues for future research.

## 2 Literature Review

Historically, the role of technology in education has evolved from passive media, such as radio and television, to interactive platforms, including e-learning modules and virtual classrooms. AI, especially chatbots, represents the next phase, where learning is not only interactive but also adaptive. With recent advances in AI technology, research interest in AI use for education has surged since 2021 (Compton & Burke, 2023; Xu & Ouyang, 2022). The public release of ChatGPT in November 2022 further heightened this interest, given the potential of the tool for educational applications.

Much of the literature addresses ChatGPT's potential benefits and pitfalls, mostly from a theoretical standpoint. ChatGPT is believed to facilitate personalized learning experiences. It provides immediate feedback, fosters self-paced learning, and immerses students in interactive learning experiences (AlAfnan, 2023; Baidoo-Anu, 2023; Kasneci et al., 2023). Recognizing the diversity within classrooms, the ability of ChatGPT to tailor responses to individual queries is a notable advantage. It aids learners in building knowledge and understanding by offering explanations, addressing questions, fostering dialogs, providing examples, and elucidating topic relevance. This interactive approach encourages students to post questions they might otherwise withhold in a classroom setting, receive meaningful feedback, and revisit previous interactions, thus promoting continuous learning.

However, the introduction of ChatGPT to the educational landscape is not without challenges. Technical limitations, such as text-only input and output constraints, currently inhibit its ability to process visual content such as images or diagrams. However, plugins such as "Show Me" have already begun to bridge this gap. Notably, the basic version of ChatGPT only encompasses data up to September 2021, but extensions like "WebPilot" allow it to search the internet for updated information. Data security remains a concern, with user prompts potentially informing future model training. Another significant challenge is the potential role of ChatGPT in academic dishonesty, such as completing student assignments. This underscores the need to re-evaluate homework and assessment design rather than question the educational utility of ChatGPT (Adiguzel et al., 2023). Overreliance on AI is another concern. Although AI can amplify learning experiences and offer personalized support, it should not overshadow the irreplaceable role of educators or the value of human interaction in learning environments. Educators provide context, promote critical thinking, and nurture socioemotional learning (Farrokhinia et al., 2023). It is paramount to recognize that ChatGPT, despite appearing knowledgeable, does not genuinely "understand" user prompts or their responses. Deriving its knowledge from vast textual data, it may inadvertently propagate problematic content, leading to biased or erroneous responses.

Several studies have investigated the proficiency of ChatGPT in standardized tests across fields such as medicine (Gilson et al., 2023), mathematics (Frieder et al., 2023), physics (West, 2023), law (Choi et al., 2023), and economics (Geerling et al., 2023). Most of these studies have reported positive outcomes.

Several studies emphasize the nuances of learning processes. Within these contexts, it is vital to distinguish between overt errors—deriving from ChatGPT's acknowledged inability to address specific questions or resulting from misinterpretations—and subtle mistakes that require in-depth subject knowledge to identify. While the former may be merely inconvenient, the latter has the potential to foster misconceptions, as a result adversely affecting learning outcomes. Gregorcic and Pendrill (2023) scrutinized the utility of ChatGPT (version 3.5) in supporting physics learning through a case study, deeming it unsatisfactory. Conversely, using GPT 4, Cooper (2023) reported more encouraging outcomes in physics, although acknowledging occasional inaccuracies. Siegle (2023) assessed its efficacy for gifted students, yielding positive findings. Conversely, Frieder et al. (2023) observed that while GPT 4 proves beneficial for mathematics up to the undergraduate level, it falters with advanced problems. Corroborating this, Wardat et al. (2023), after interviewing 30 individuals experienced in using ChatGPT for mathematics, reported analogous findings. Arndt (2023) found ChatGPT to be a valuable assistant for Systems Thinking, though noted occasional overt errors and subtler logical inconsistencies. Interestingly, these logical challenges echo the issues identified in physics. In the realm of economics, Arndt (2023) not only assessed the answers of ChatGPT to three economic tests but also evaluated the quality of its explanations. While GPT 3.5 produced approximately 5% incorrect answers capable of inducing misconceptions, its successor, GPT 4, exhibited none. In summary, GPT 4 evidently outperforms GPT 3.5, which is a predictable progression. Current research suggests that while ChatGPT usually excels in knowledge-based tasks, it occasionally fails at logical conclusions. It is also worth highlighting that the quality of responses from both ChatGPT versions can fluctuate considerably over short spans, as observed by Chen et al. (2023).

Considering the technical aspects of ChatGPT is essential for its educational integration, but user perceptions and acceptance are equally crucial. Biggs (2011) introduced a comprehensive framework, the 3P model, emphasizing the interplay between 'Presage', 'Process', and 'Product' in the educational sphere. This model suggests that students' perceptions of their learning environment heavily influence their approach and outcomes. Positive perceptions drive deep comprehension, while negative ones lean toward rote memorization. In this context, students' perceptions of AI tools such as ChatGPT can either facilitate or impede their integration into the learning process.

Previous studies have demonstrated that AI tools can elevate student performance, self-confidence, motivation, and attitude. Students typically value their responsiveness, interactivity, and discreet support (Bailey et al., 2021; Essel et al., 2022; Gayed et al., 2022; Lee et al., 2022). Factors influencing students' attitudes toward AI tools include computer self-efficacy, self-directed learning, trust in the tool, and a sense of social presence (Choi et al., 2023), as well as perceived utility and ease of use (Hew et al., 2023).

Given the short time since its release, empirical studies specifically addressing ChatGPT are still scarce. Mogavi et al. (2023) conducted a content analysis of four major social media platforms and identified widely varying opinions of the usefulness of ChatGPT for educational purposes. A study conducted by Chan and Hu (2023) where they asked 399 university students about generative AI such as ChatGPT. The participants acknowledged the possibilities for individualized learning aid, assistance in writing and idea generation, and research and analytical skills. However, they also voiced apprehensions regarding correctness, confidentiality, ethical dilemmas, and the effects on personal growth.

## 3. Research Questions

The evident potential of chatbots, particularly ChatGPT, in education underscores the need for empirical studies evaluating their efficacy as learning tools. Furthermore, comprehending learners' perceptions, concerns, and experiences is vital to seamlessly integrate these tools into mainstream educational settings. Consequently, this research posits the following questions:

RQ1: How do users perceive chatbots, specifically ChatGPT, prior to exposure? This involves exploring initial concerns and expectations and analyzing any correlations with variables such as gender or previous chatbot experiences.

RQ2: How do these perceptions evolve postexposure? This will delve into potential shifts in attitudes and examine any correlations with pertinent factors.

RQ3: What challenges and benefits do users experience while learning with ChatGPT, and how can their learning journey be further enhanced?

## 4. Methodology

To delve into the role of ChatGPT within educational contexts, a mixed-methods approach was adopted, combining quantitative and qualitative research techniques. This combination afforded a richer, multifaceted insight into the influence of ChatGPT and the experiences of the participants.

**Participants**:
A diverse group was selected for the study. Participants provided demographic details, such as age, gender, occupation, and previous encounters with chatbots.

**Study Design:**

The research was structured around a pretest-posttest model incorporating the following components:

1. Pre-Exposure Questionnaire: This initial survey captured participants' demographics, prior chatbot interactions, perceptions, apprehensions, and expectations regarding chatbot-assisted learning. Additionally, participants selected up to three economic or business subjects they wished to explore, rating their preexisting knowledge and interest in these areas.

2. ChatGPT Interactive Session: Participants engaged with ChatGPT to learn about their chosen subjects. Initially, they received a set of guidelines, particularly emphasizing effective prompting strategies. An expert proficient in both ChatGPT-aided learning and the fields of economics and business observed these sessions. Their observations centered on the progress in learning, challenges faced, content accuracy, and the overall learning environment. The experts also offered assistance when needed or requested.

3. Post-Exposure Questionnaire: Following the ChatGPT session, participants completed another survey, mirroring the initial one in structure. They were also prompted to reflect on their learning journey, outcomes, and perceptions of ChatGPT as an educational tool.

Both surveys incorporated open-ended questions and five-point Likert scales, with "1" denoting disagreement and "5" indicating full agreement. The Likert scales, with minor modifications, encompassed the following:

- Competence - I feel competent in dealing with chatbots.
- Insecurity - Interacting with AI and chatbots causes insecurity or fear in me.
- Understandability - I understand the statements made by chatbots well.
- Quality of Responses - The quality of the responses is good.
- Fun-general - I enjoy interacting with chatbots.
- Fun-learning - I enjoy learning with chatbots.
- Improve learning - Interacting with chatbots has the potential to improve my learning.

- Usage in class - I wish chatbots were used in classes (or I would have wished it for my earlier school years).
- Curiosity - I am curious about how to interact with chatbots.
- Feel comfortable - I feel comfortable interacting with chatbots.
- Often used for learning - I have often worked with chatbots to learn something or discover something new.

For each subject the participants sought to explore, the initial survey captured their preexisting knowledge and interest. Postinteraction, they rated their comprehension, and the helpfulness and quality of ChatGPT's assistance.

## 5. Results

### 5.1. Demographics

Table 1 illustrates the age distribution of the participants and their prior experience with chatbots. The average age was 22.8 years with a standard deviation of 13.8, spanning from 8 to 79 years. A significant correlation (Cramer's V = 0.566, p < 0.001) exists between age and previous experience with chatbots, with both the youngest and oldest cohorts showing limited exposure. Out of the total, 60 participants were school students, 22 were university students, and 19 were employed.

| Age Range | Total n | With experience | Without experience |
| --- | --- | --- | --- |
| 8-12 | 12 | 1 | 11 |
| 13-18 | 41 | 35 | 6 |
| 19-25 | 30 | 23 | 7 |
| >26 | 19 | 7 | 12 |
| Total | 102 | 66 | 36 |

Table 1: Age distribution and prior chatbot experience

The gender breakdown reveals a slight female majority (n=53) compared to males (n=44). One participant identified as diverse in gender, while four chose not to specify. Notably, 70% of male respondents have prior chatbot experience, in contrast to 56% of females. However, a chi-square test found no significant correlation between gender and prior experience ($p > 0.05$).

**5.2. Pretreatment perceptions about learning with chatbots (RQ1)**

Table 2 outlines the perceptions of the 66 participants with prior chatbot experience. Generally, these participants rated most aspects favorably, reflecting a positive attitude toward chatbots, despite limited experience in learning by use of this medium (M = 2.78). The 36 participants without prior experience, who answered hypothetically, showed moderate opinions about learning with chatbots. However, they exhibit strong curiosity about the technology (M = 4.09, SD = 0.8).

|  | With experience Mean | SD | Without experience Mean | SD | After treatment Mean | SD | Weighted difference Δ-Mean |
|---|---|---|---|---|---|---|---|
| Competence | 3.44 | 1.025 | 3.24 | 0.936 | 4.11 | 0.884 | 0.74 |
| Insecurity | 1.95 | 1.082 | 2.58 | 1.339 | 1.73 | 1.043 | -0.44 |
| Understandability | 4.2 | 0.775 | 3.25 | 1.105 | 4.16 | 0.918 | 0.30 |
| Quality of Responses | 3.74 | 1.02 | 3.36 | 1.046 | 4.13 | 0.787 | 0.52 |
| Fun-general | 3.94 | 0.95 | 3.33 | 1.069 | 4.15 | 0.925 | 0.43 |
| Fun-learning | 3.35 | 1.165 | 3.28 | 1.059 | 3.99 | 1.000 | 0.66 |
| Improve learning | 3.63 | 1.162 | 3.17 | 1.207 | 4.02 | 0.979 | 0.55 |
| Usage in class | 3.74 | 1.241 | 2.94 | 1.351 | 3.91 | 1.173 | 0.45 |
| Curiosity |  |  | 4.09 | 0.843 |  |  |  |
| Feel comfortable | 3.62 | 1.056 |  |  | 3.99 | 0.995 | 0.37 |
| Often used for learning | 2.78 | 1.303 |  |  |  |  |  |

Table 2: Participants' perceptions of chatbots before and after treatment

In the preliminary questionnaire, prior to the learning session, participants expressed their concerns and expectations about learning with chatbots. Forty-three participants indicated that they were concerned about the accuracy of the information provided. This is closely linked to the desire among participants to verify the information provided by the chatbots and determine the sources of that information. A lack of transparency was recurring criticism, mentioned 15 times. Some participants (6) felt that chatbot responses might not align with their specific needs, being too generic, superficial, confusing, challenging to understand, or complex. Data privacy and the storage of their prompts/inputs was a concern for 11 participants. A smaller group (4) worried about the blurring lines between human and machine interactions and the potential impact on human relationships. There was also a notable fear (18 mentions) that overdependence on chatbots could diminish cognitive abilities or hinder independent thinking. Interestingly, only two participants highlighted the potential challenge of formulating effective prompts for the chatbot.

Parallel to concerns about misinformation, there was a strong expectation (voiced 15 times) that chatbots should provide clear and accurate answers. The most frequent expectation (30 mentions) was that chatbots would enhance the speed and efficiency of information retrieval. ChatGPT was widely seen (18 mentions) as a tool that could support general learning processes. Its potential to facilitate interactive and personalized learning was recognized seven times. Twelve participants anticipated it would assist in knowledge acquisition, while four believed it could boost higher cognitive skills such as creativity, critical thinking, or media literacy. The prospect of a more engaging and enjoyable learning experience with ChatGPT was expressed four times.

### 5.3 Posttreatment Perceptions about Learning with Chatbots (RQ2)

On average, inclusive of a brief introduction and the time taken to complete questionnaires, sessions lasted 52 minutes (SD = 17). Overall, participants delved into 230 different topics. On average, participants spent 17.5 minutes (SD = 10.4) and utilized 7.5 prompts (SD = 4.6) per topic. Postlearning, participants reported a high level of understanding (M = 4.1, SD = 0.9) and found the responses of ChatGPT to be helpful (M = 4.3, SD = 0.9) and of high quality (M = 4.2, SD = 1.0).

Post-session ratings, as illustrated in table 2 columns 5 and 6, showed a marked improvement in most areas, indicating that the learning experience exceeded participants' expectations. The most significant upticks were observed in competence using chatbots (+0.74) and the enjoyment derived from learning with chatbots (+0.66).

Post-session data also revealed that those with prior chatbot experience showed marginally higher ratings in competence (+0.26) and general enjoyment with chatbots (+0.28). However, these differences were not statistically significant. Other aspects showed only trivial variations, suggesting that posttreatment attitudes toward ChatGPT became largely independent of prior chatbot experience.

Gender-based analysis showed that males generally reported higher values across aspects (except insecurity) compared to females. Notably, a significant difference was observed in the 'improve learning' perception. Males (M = 4.31, SD = 0.7) scored higher than females (M = 3.83, SD = 1.1); $t(92) = 2.458$, $p = 0.02$, indicating a large effect size (Cohen's $d = 0.95$).

Participants who used ChatGPT V4 consistently reported higher scores in various aspects compared to those who engaged with V3.5. Notable differences were observed in categories such as general enjoyment ($d = 0.908$, $p = 0.010$), feeling of insecurity ($d = 1.053$, $p = 0.018$), comfort during interaction ($d = 0.977$, $p = 0.012$), clarity of responses ($d = 0.897$, $p = 0.018$), response quality ($d = 0.757$, $p = 0.032$), enjoyment derived from learning ($d = 0.951$, $p < 0.001$), preference for chatbot integration in classrooms ($d = 1.121$, $p = 0.006$), and the perceived potential for chatbots to enhance learning ($d = 0.936$, $p = 0.002$). These statistics highlight a sentiment among participants that GPT 4 offers a markedly improved experience compared to GPT 3.5.

Table 3 offers insights into the correlations between various variables. Participants without prior chatbot experience were driven by their inherent curiosity. For them this was the primary determinant of satisfaction with chatbot-based learning. For participants without prior chatbot experience, the primary determinant of satisfaction with chatbot-based learning was their inherent curiosity. This trait exhibited moderate correlations with posttreatment factors such as perceived competence, comfort during interactions, and the potential of chatbots to enhance learning.

Post-session, participants' perceived competence with chatbots. This correlates moderately with their comprehension of the discussed topic and their assessment regarding the helpfulness of ChatGPT's responses.

Furthermore, participants' belief in ChatGPT's ability to support their learning was strongly associated with emotional factors. These included enjoying the interaction, finding learning with the tool pleasurable, and feeling at ease during the engagement. Strong correlations were also found with the participants' self-assessed competence, clarity and helpfulness of the responses of ChatGPT and a desire for chatbot integration in educational settings. When exploring participants' comprehension of specific subjects, similar but moderately strong correlations were observed.

Last, the desire to incorporate ChatGPT into classroom settings was highly correlated with its perceived capacity to enhance learning and was closely linked to positive emotional responses, particularly the enjoyment derived from both the interaction and the learning process.

| Variable 1 | Variable 2 | r | p |
|---|---|---|---|
| Curiosity (pre) | Competence (post) | .46 | .002 |
| | Fun general (post) | .35 | .02 |
| | Feel comfortable post) | .43 | .005 |
| | Improve learning (post) | .31 | .04 |
| | Topic helpful | .30 | .04 |
| Competence (post) | Topic understanding | .31 | .002 |
| | Topic helpful | .36 | <.001 |
| Improve learning (post) | Competence (post) | .49 | <.001 |
| | Fun general (post) | .65 | <.001 |
| | Fun-learning | .71 | <.001 |
| | Feel comfortable post) | .62 | <.001 |
| | Understandability (post) | .57 | <.001 |
| | Answer quality (post) | .56 | <.001 |
| | Usage in class | .68 | <.001 |
| | Topic helpful | .54 | <.001 |
| Topic understanding | Competence (post) | .31 | .002 |
| | Fun general (post) | .33 | .001 |
| Usage in class (post) | Understandability (post) | .38 | <.001 |
| | Fun-learning | .41 | <.001 |
| | Usage in class | .39 | <.001 |
| | Improve learning (post) | .34 | .001 |
| | Improve learning (post) | .68 | <.001 |
| | Competence (post) | .28 | .006 |
| | Fun general (post) | .65 | <.001 |
| | Insecurity (post) | -.37 | <.001 |
| | Feel comfortable post) | .50 | <.001 |
| | Understandability (post) | .47 | <.001 |
| | Answer quality (post) | .53 | <.001 |
| | Fun-learning | .71 | <.001 |
| | Improve learning (post) | .68 | <.001 |
| | Topic understanding | .39 | <.001 |
| | Topic helpful | .41 | <.001 |

Table 3: Selected correlations

## 5.4. Issues, Positive Experiences, and Enhancement Recommendations (RQ3)

Following their learning experience with ChatGPT, participants shared feedback regarding perceived challenges, suggestions for improvement, positive aspects of their experience, and other relevant observations.

Before exposure, the accuracy of the provided answers was a primary concern. After exposure, this concern was only mentioned five times, with participants noting confusing responses or feeling that they were misunderstood. Trust in the replies of ChatGPT was another frequently voiced issue, with nine participants expressing reservations. While three participants found the answers of the chatbot to be overly general, a significant number (25 mentions) felt that the responses were excessively verbose or complex. Although concise and precise prompts can mitigate this, 21 participants found phrasing their prompts challenging. Other feedback touched upon the lack of current information (five mentions) and the absence of visual aids in the answers (also five mentions). Three participants were not satisfied with the quality of ChatGPT's assessment, questions or the feedback they received in response.

On the technical side, six participants encountered issues such as delayed responses, nonsensical answers, or complete nonresponsiveness, issues that were typically resolved with a system restart. Four participants maxed out their prompt allocation, which made it necessary to switch to the previous version, GPT 3.5.

Experimenters' feedback largely mirrored the participants' sentiments. They observed common challenges such as a lack of up-to-date information, participants struggling with phrasing prompts, verbose responses, and a lack of visual aids. Additionally, experimenters noted instances where users ventured into unrelated topics due to an apparent lack of basic knowledge. When assessment tests were employed, answers were occasionally provided alongside questions. A few instances of network errors were also reported.

In an exploration of 230 distinct topics, five errors were detected that had the potential to lead to misconceptions. These errors pertained to types of goods, economic principles, and economies of scope.

Regarding improvements, 12 participants expressed a desire for references or sources, aligning with concerns about the accuracy of the answers. In line with feedback about verbose responses, 13 participants hoped for more concise and straightforward answers. Seven emphasized the importance of up-to-date information, while five suggested voice input capabilities. A significant number (12 participants) felt that visual aids would enhance comprehension, especially for complex subjects. Furthermore, five participants voiced a need for guidance, particularly in phrasing prompts, whether from ChatGPT itself or external entities such as educators or peers.

Regarding positive experiences, a significant majority (77 participants) appreciated the explanations provided by ChatGPT, noting the ease and speed with which they received helpful information. Twenty participants valued the interactive and personalized learning approach of the platform. Five praised the examples given.  Although many participants did not take advantage of ChatGPT's capability to assess their knowledge, ten found this feature and the feedback beneficial. Five participants explicitly stated their enjoyment of using ChatGPT for learning.

Experimenters typically characterized the learning atmosphere with descriptions such as "interested," "curious," "focused," and "fascinated." Initial skepticism in some participants usually shifted to interest, although three remained skeptical throughout. Four participants displayed only mild interest. One participant, who began with skepticism, encountered technical issues, leading to increased frustration and skepticism. Generally, the positive learning atmosphere persisted when studying a second topic but decreased when participants delved into a third topic, as weariness and losing focus became evident.

Regarding additional remarks, some criticisms emerged. Four participants felt that learning with ChatGPT was either boring or inferior to traditional methods, and concerns about increased screen time and data security were raised. Two participants feared an overreliance on AI for learning, while eight acknowledged the utility of ChatGPT but emphasized a critical approach to its responses. The sentiment that ChatGPT cannot replace the social aspects of learning was mentioned twice. Six participants anticipated advancements in AI learning tools, eleven already found ChatGPT beneficial, and nine enjoyed their experience. Three highlighted the importance of effectively using ChatGPT.

Some participants felt uncertain about their comprehensive understanding of topics, pointing to a potential gap in overarching knowledge. Most had prior experiences with GPT 3.5 and expressed surprise at the enhanced learning experience with GPT 4. Interestingly, one participant, despite praising ChatGPT, was against its integration into formal education, fearing school would ruin the experience. This view inadvertently exemplified the overjustification effect, where intrinsic motivation diminishes in the presence of external rewards (Carlson & Heth, 2007).

Experimenters frequently emphasized the crucial role of prompt phrasing. The quality of responses, and consequently user satisfaction and learning outcomes, hinged on effective prompts. Despite offering prompt phrasing guidelines and in-session advice, many participants chose to navigate the experience independently, often improving their interactions based on prior encounters.

# 6. Discussion

## 6.1. Findings

### 6.1.1. Leveling competence and perceptions

The initial phase of this study aimed to understand users' pretreatment perceptions about chatbots. This foundational exploration revealed an interplay between prior experience with chatbots and initial perceptions. Those without prior exposure to chatbots entered with average expectations. In contrast, participants with prior experience showed above-average perceptions. This implies that informal learning about using chatbots by simply using them leads to positive perceptions, including competence and knowledge of how to use them for learning. However, posttreatment data revealed a notable improvement in all perceptions, especially in perceived competence. It is interesting to note that no significant differences between the groups could be identified anymore. This suggests that even a short structured and guided session can level differences and implies that formal training, e.g., in schools, on how to use chatbots for learning might be a superior way to learn how to learn with chatbots.

### 6.1.2. The Evolving Trust in AI-Driven Answers

Before engaging with ChatGPT, participants predominantly expressed apprehensions about the accuracy of the responses of the platform. Posttreatment data showed a substantial reduction in these concerns, likely attributed to the mere detection of five potential misconceptions. This change highlights the evolving trust in the capabilities of ChatGPT, although some participants still point out limited trust in its answers and advise maintaining a critical mindset. Pedagogically, this general skepticism fostering critical thinking is a cornerstone of effective learning, echoing principles from Bloom's taxonomy (Bloom, 1953; Athanassiou et al., 2003). With AI tools, this skepticism becomes especially vital, given the nuanced risks of misinformation.

For learners embarking on autodidactic journeys, this poses a distinctive challenge. They typically rely on AI tools for areas outside their expertise, making it difficult to discern inaccuracies. A holistic approach to mitigate this issue would involve cross-referencing the responses of ChatGPT, although this can be time intensive. Consequently, nurturing the ability to gauge the credibility of AI-generated information emerges as a pivotal facet of an AI-learning competence.

### 6.1.3. Sourcing, Verification, and Changing Perceptions

The desire for sourced information remained consistent pre- and posttreatment. While ChatGPT can reference scientific sources, a word of caution is in order: these sources make meticulous verification necessary due to comparatively frequent inaccuracies or fabrications.

### 6.1.4. The Challenge of Prompting

A remarkable revelation was the underestimation of the importance of prompting. Pretreatment, prompting was mentioned only twice as a potential challenge. This number rose significantly posttreatment, although participants received guidance on effective prompting. This indicates that prompting is a more profound challenge than anticipated. Effective prompts not only elicit precise responses but also adapt explanations to individual preferences, which would reduce the often-mentioned complaint about long or hard-to-understand explanations.

### 6.1.5. Toward a Holistic AI-Learning Competence

The importance of AI-learning competence is supported by the findings that perceived competence correlates with the understanding of topics and finding the answers of ChatGPT helpful for learning. Based on the results of the study, first central elements of an AI learning competence can be identified:

- Critical thinking with respect to the accuracy of the information given. Learners should be able to assess the credibility of the information based on appropriate criteria and estimate when it is appropriate to cross-check the given information.
- The ability to craft effective prompts.
- Knowledge about data security.
- The ability to choose the most fitting AI tool for a specific task based on knowledge about their capabilities and limitations. For ChatGPT, this would include its ability to assess users' understanding by asking them questions, as well as being familiar with the newly introduced 'Code Interpreter' that greatly expands the capabilities of the tool. Moreover, it is helpful to know the abilities of plugins that, among other things, enable visualizations—a skill many participants wished for.
- Core self-learning competencies such as knowing how to obtain an overview of new topics, how deep to delve into these topics, how not to digress into irrelevant details, and how to test and improve learning progression.

There are established theoretical frameworks with a focus on technology and digital competences for educators, such as TPACK (Mishra & Koehler, 2006) and DigCompEdu (Redecker & Punie, 2017), as well as frameworks for learners, such as P21 Century Skills (Partnership for 21st Century Learning, 2015) and DigComp (Vuorikari et al., 2022). However, given their general character, they do not specifically address competencies necessary for making use of AI in learning processes. Given their importance, it seems appropriate to develop a specific AI-learning competence framework that might include some of the abovementioned suggestions.

### 6.1.6. Integrating AI in Formal Education

To bridge the looming digital divide, AI should be integrated into formal educational settings. First and foremost, this would enable all students to enhance their AI-learning competence. Furthermore, tools such as ChatGPT provide opportunities for adaptive, individualized, and highly interactive learning and information retrieval. These tools can enhance the quality of learning and teaching, especially when using the advanced version, GPT-4, instead of GPT 3.5. Such an approach aligns closely with constructivist learning theories.

In class, teachers with didactic expertise can integrate AI tools in a way that successfully addresses concerns participants have mentioned, such as creating misconceptions due to inaccurate information, reduced social learning, or becoming dependent on the technology. Furthermore, the adaptability and interactivity of tools such as ChatGPT offer a paradigm shift in pedagogy. The positive inclination toward AI, even among initially neutral or skeptical users, further accentuates its potential.

## 6.2 Limitations

The study has provided insights into the use of ChatGPT as an educational tool. However, it is essential to recognize its limitations to ensure a balanced interpretation. First, the sample size, although diverse, might not be expansive enough to allow for broad generalizations. This limitation is exacerbated by the potential underrepresentation of specific demographic groups. In particular, there seems to be a void in regard to younger learners such as elementary school students, a group that could offer a fresh perspective on AI tool utility.

Another consideration is the exclusive focus on ChatGPT. While ChatGPT is certainly a leader in the AI chatbot realm, the vast landscape of AI-driven educational tools means that other chatbots could offer differing experiences that were not captured in this study. This exclusivity also means that the findings are specifically tied to ChatGPT's functionalities and limitations.

The temporal scope of the study is another constraint. The reliance on short-term engagements with ChatGPT does not provide insights into potential long-term effects or changing usage patterns over time. Extended interactions might reveal evolving perceptions, emergent challenges, or even new benefits.

Moreover, the methodology of the study, which leans heavily on self-reported data, could be a double-edged sword. Even though participants' responses were supplemented with experimenter comments, the possibility of bias cannot be entirely dismissed. Participants could be influenced by a myriad of factors, from wanting to meet the researcher's expectations to misunderstanding their learning outcomes. An objective evaluation, perhaps through a pre- and posttreatment knowledge test, would offer a more concrete assessment of learning efficacy.

Last, the subject specificity of the study — focusing on business and economics — may not reflect the performance of ChatGPT across various academic fields. Other disciplines, especially those that necessitate intricate logical reasoning, could pose more significant challenges for ChatGPT, potentially leading to varying user experiences and perceptions.

**6.3 Future Research**

Given the nascent stage of AI integration into education and the limitations identified, there is an immense scope for future research. A pressing area to delve into is the accuracy assessment of AI tools. Identifying and testing robust criteria to evaluate the veracity of AI-generated responses is paramount, especially in an educational setting. Alongside this, strategizing ways to navigate the challenges of misinformation or inaccuracies is crucial.

The emergence of the concept of 'AI-learning competence' from this study suggests a broader educational paradigm shift. However, before this construct gains widespread acceptance, it requires rigorous theoretical modeling. This includes the creation of a specific framework and potential adaptation of existing ones.

Longitudinal studies could bridge the temporal limitations of the current studies by tracking AI tool interactions over prolonged periods. Such studies would offer insights into evolving user perceptions, usage patterns, and potential long-term effects or benefits.

Last, given the linchpin role teachers play in classroom learning, understanding their perceptions and competencies regarding AI is essential. This could pave the way for tailored teacher training modules that not only equip educators with the skills to use AI tools effectively but also empower them to guide their students in navigating the AI-augmented educational landscape.